# Some Critical Support for Power Laws and their variations

Pedro Miramontes [1,4,*] Wentian Li [2] and Germinal Cocho [3,4]

1. Departamento de Matemáticas, Facultad de Ciencias, Universidad Nacional Autónoma de México, México 04510 DF, México.
2. The Robert S. Boas Center for Genomics and Human Genetics, The Feinstein Institute for Medical Research North Shore LIJ Health System, Manhasset, 350 Community Drive, NY 11030, USA.
3. Instituto de Física. Universidad Nacional Autónoma de México, México 04510 DF, México.
4. Centro de Ciencias de la Complejidad. Universidad Nacional Autónoma de México, México 04510 DF, México.

In "Critical Truths About Power Laws" (*Science*, **335**, pp665-666), Stumpf and Porter, give us a critical view about the widespread usage of power laws to model some phenomena or simply to organize data with acceptable goodness of fit. However, they are only partially right when state that "most reported power laws lack statistical support and mechanistic backing" as they do not take into account of all "historical" power laws in science such as, just to mention a few, Coulomb's inverse square law of electrostatic force, Kepler's laws of planetary motion or Stefan-Boltzmann law of radiation. Admittedly, these are not statistical distributions, but undeniably, these power-laws enjoy support from data as well as mechanistic backing.

A power-law probability density function can be converted to a power-law rank-size function, with the "long tail" becoming the "head". One of the best known examples of this form of representation of a power-law is the Zip's law of word usage in a language. With little effort, we can show that power laws, like Zip's law, can be immersed in a more general, two parameter, family of functions (1):

$$f(r) = A\frac{(N-r)^b}{r^a}$$

where *r* is the rank, *N* the number of cases and *A* a proportionality constant. The power-law rank function is recovered by b → 0. This new family of functions is called beta function or discrete generalized beta distribution (DGBD). For many datasets, where a power-law distribution fails to fit well, it could nevertheless be fit perfectly by DGBD. DGBD also emerges as a robust distribution under very general conditions. Here we will provide some evidence supporting our claim that DGBD, while containing power laws as a special case, has a strong statistical support and mechanistic sophistication.

Several studied examples ranging from the rank-order distribution of impact factor of journals to the size of geometric patterns in abstract painting can be found in (3). The goodness of the fit is astonishing. An exhaustive comparison of DGBD to other two-parameter models, including the Yule distribution, is presented in (4) for several statistical linguistic data, and DGBD is often the best fitting function. This invalid an argument that DGBD could fit data better because it contains one extra parameter, as DGBD also outperforms other functions with the same number of parameters. Besides enjoying statistical support, more importantly, we believe DGBD also possesses what Stumpf and Porter called "mechanistic sophistication".

* Corresponding autor. E-mail: pmv@ciencias.unam.mx

One possible mechanism behind a DGBD distribution is the expansion-modification model (5). This is a dynamical system equivalent to a coupling of discrete reactive and diffusive mechanisms. Ranked frequencies of substring types generated by the model are shown to follow the DGBD, and the fitting parameter values are intrinsically related to the parameter in the model which controls the order-disorder transition (6). In another classic example, as shown by Kolmogorov, the distribution of the energy of vortices, as a function of their size in turbulence studies, distribute as a short tail power law (7). The power-law fitting turns out not to be perfect, but a very high goodness fitting with a DGDB is observed. In the case of allometry, being near to a power-law, not surprisingly one gets a DGDB with $b\sim 0$. On the other hand, when one considers interacting bodies with stochastic potentials (attractive as well as repulsive), the difference of the diagonal matrix energy values distributes as a DGDB with $a\sim 0$ and $b$ close to 1 (8)

The list can be continued, but we think that the point is made that a modified power laws, a more general class of functions which include power-law as a special case, have good statistical supports and mechanistic backing. The work on DGBD is ongoing, and there are more datasets to check and more fundamental questions can be asked. However, we believe the extremely good fitting to DGBD of many real datasets indicates a universal appeal of the model, and it should find more applications in economics, ecology or molecular biology on data that wait for a proper explanation.

1. Stumpf MPH and Porter MA. *Science.* **335**, 665 (2012).
2. Mansilla R, Köppen E, Cocho G, and P. Miramontes. J. *Informetrics*. **1**, 155 (2007).
3. Martínez-Mekler G, Alvarez Martínez R, Beltrán del Río M, Mansilla R, Miramontes P, and G. Cocho. *PLoS ONE* **4**(3), e4791. doi:10.1371/journal.pone.0004791. (2009).
4. Li W, Miramontes P and Cocho G. *Entropy.* **12**, 1743 (2010).
5. Li, W. *Physical Review* E. **43**, 5240 (1991).
6. Alvarez-Martinez R, Martinez-Mekler G, Cocho G. *Physica* A, **390** (2011)
7. Saddoughi SG (1997). *J. Fluid Mech*. **348**, 201 (1997).
8. Hernández-Quiroz S, Beltrán M, Benet L, Flores J, and G. Cocho. http://arxiv.org/abs/1203.6694. (2012).